\documentclass{article}
\usepackage{spconf,amsmath,graphicx,hyperref}
\usepackage{algorithm}
\usepackage[noend]{algpseudocode}
\usepackage{float}

\usepackage{amssymb}
\usepackage{booktabs}
\usepackage{arydshln}
\usepackage{enumitem}
\usepackage{epstopdf}
\usepackage[font=small,skip=0pt]{caption}
\title{Int-MeanFlow: Few-step Speech Generation with Integral Velocity Distillation}
\name{Anonymous Submission}
\address{}

\name{Wei Wang, Rong Cao, Yi Guo,  Zhengyang Chen, Kuan Chen, Yuanyuan Huo}
\address{ByteDance}
\begin{document}
\ninept
\maketitle
\begin{abstract}

Flow-based generative models have greatly improved text-to-speech (TTS) synthesis quality, but inference speed remains limited by the iterative sampling process and multiple function evaluations (NFE). The recent MeanFlow model accelerates generation by modeling average velocity instead of instantaneous velocity. However, its direct application to TTS encounters challenges, including GPU memory overhead from Jacobian-vector products (JVP) and training instability due to self-bootstrap processes.
To address these issues, we introduce IntMeanFlow, a framework for few-step speech generation with integral velocity distillation. By approximating average velocity with the teacher’s instantaneous velocity over a temporal interval, IntMeanFlow eliminates the need for JVPs and self-bootstrap, improving stability and reducing GPU memory usage. We also propose the Optimal Step Sampling Search (O3S) algorithm, which identifies the model-specific optimal sampling steps, improving speech synthesis without additional inference overhead. Experiments show that IntMeanFlow achieves 1-NFE inference for token-to-spectrogram and 3-NFE for text-to-spectrogram tasks while maintaining high-quality synthesis. Demo samples are available\footnote{\url{https://vvwangvv.github.io/intmeanflow/}}.

\end{abstract}
\begin{keywords}
Text-to-speech, flow matching, meanflow, efficiency, speed-quality tradeoff
\end{keywords}
\vspace{-.5em}
\section{Introduction}
\label{sec:intro}

Text-to-speech (TTS) generation has made significant progress in recent years, with models achieving near-human-level, zero-shot synthesis capabilities~\cite{cosyvoice2-du2024, cosyvoice3-du2025, F5TTS-chen2024, xtts-casanova2024, yourtts-casanova2022, seedtts-anastassiou2024, zipvoice-zhu2025}. The rise of flow-based generative models~\cite{fm-lipman2022} has contributed to this advancement, offering promising results across various fields, including image synthesis~\cite{Flow-albergo2023, RF-liu2022}, video~\cite{VideoFlow-Jin2024} and music generation~\cite{MusicFlow-KR2024}. 
These models learn to map data distributions to a latent space, enabling high-quality generation. However, flow-based models often face a trade-off between sampling quality and efficiency, as their iterative sampling process can lead to slow inference and high computational costs. To address this, recent efforts in the image domain, such as consistency models~\cite{audiolcm-lu2024, musiccm-fei2024, cm-song2023, simplifyingcm-lu2024, multistepcm-heek2024, facm-peng2025} and shortcut models~\cite{sm-kevin2025}, have been proposed to reduce the number of function evaluations~(NFE) while maintaining high-quality generation results. 

Among these approaches, MeanFlow~\cite{MeanFlow-geng2025} has emerged as a promising solution to the trade-off between sampling quality and efficiency. By modeling averaged velocity instead of instantaneous velocity, MeanFlow enables more efficient sampling without compromising output quality. While it has shown success in image generation and audio generation~\cite{MeanAudio-li2025}, applying MeanFlow to TTS introduces several challenges. First, the training process of MeanFlow relies on a self-bootstrap mechanism, which requires mixing with instantaneous velocity guidance similar to flow matching. The strength of this guidance can significantly impact model performance, and without it, the model is prone to collapse. Second, MeanFlow involves the Jacobian-vector product (JVP), a computationally intensive operation that consumes substantial GPU memory. Additionally, JVP is not natively supported by certain custom CUDA operators or torch-native operations, such as flash attention, complicating the adaptation of existing models to the MeanFlow framework. These memory and compatibility challenges make training large-scale TTS models with MeanFlow infeasible.

To address these challenges, we propose IntMeanFlow, a framework for few-step speech generation that leverages integral velocity distillation. Building on the core motivation of MeanFlow, IntMeanFlow enables the model to learn averaged velocity instead of instantaneous velocity. Motivated by the observation that the quality of speech generated by flow-based models plateaus after a certain NFE, we approximate the average velocity over a temporal interval $[t,r]$ by dividing the accumulated displacement between discrete time steps by the interval duration. We further introduce an initialization strategy to leverage pretrained flow-matching models, enabling smoother migration from existing models. By eliminating the need for self-bootstrap and reliance on the Jacobian-vector product (JVP), IntMeanFlow ensures improved stability, reduced GPU memory consumption during training, and better compatibility with existing models, simplifying the adaptation process.

Additionally, based on the empirical observation that denser sampling near noisier timesteps leads to better generation quality, we introduce the Optimal Step Sampling Search (O3S) algorithm. O3S automatically identifies model-specific near-optimal sampling steps, using a customizable quality metric and a ternary search algorithm. This improves the generation quality without introducing additional inference overhead.

We conduct experiments on models from two widely used approaches in flow-based TTS: (1) CosyVoice2~\cite{cosyvoice2-du2024}, which integrates a language model (LM) followed by a flow model converting time-aligned tokens into mel-spectrograms (token2mel), and (2) F5-TTS~\cite{F5TTS-chen2024}, where a flow model directly converts raw text embeddings into mel-spectrograms (text2mel), learning time alignment inherently. Experimental results show that IntMeanFlow achieves 1-NFE inference for the token2mel task in CosyVoice2 and 3-NFE for the text2mel task in F5-TTS, maintaining high-quality synthesis.

Our contributions can be summarized as follows:
\begin{enumerate}[topsep=0pt]
    \item We propose a distillation framework that applies MeanFlow to TTS, overcoming training memory overhead and stability issues associated with its direction application to TTS.
    \item We introduce a search algorithm to identify model-specific sampling steps, improving generation efficiency without increasing inference overhead.
    \item We evaluate our method on two popular flow-based TTS models, achieving 1-NFE inference for token2mel and 3-NFE for text2mel tasks, while maintaining high-quality synthesis.
\end{enumerate}

\section{Related works}
\label{sec:related_works}

\subsection{Distillation based on Direct Metric Optimization}

Consistency and shortcut models are commonly used to reduce NFE while maintaining output quality. However, applying these methods to text-to-speech (TTS) has shown limited success due to task differences. Notable exceptions, such as DMOSpeech~\cite{dmospeech-li2024} and DMOSpeech2~\cite{dmospeech2-li2025}, reduce NFE in TTS through distillation and reinforcement learning.
However, these methods rely on auxiliary models that complicate the training process, as well as a fixed time sampling schedule during distillation, which reduces flexibility during inference. In contrast, our approach eliminates the need for auxiliary models and additional loss functions. And we do not employ a fixed step sampling strategy during training, which simplifies the training process, and improves flexibility during inference.

\subsection{MeanFlow}

MeanFlow introduces a framework for one-step generative modeling by defining average velocity as displacement over a time interval. Unlike Flow Matching, which models instantaneous velocity, MeanFlow focuses on learning average velocity and establishes an analytical relationship between average and instantaneous velocities through the MeanFlow Identity. This approach provides a clear training objective without relying on heuristic constraints, and it operates independently of score estimation, pretraining, or distillation. However, its dependence on Jacobian-vector products (JVPs) to compute the time derivative introduces computational overhead and limits scalability, especially with custom operators lacking JVP support.

\section{Methodology}
\label{sec:methodology}

\subsection{IntMeanFlow: MeanFlow Distillation via Integral Velocity}

IntMeanFlow extends the principles of MeanFlow by focusing on learning the averaged velocity over a time interval, rather than the instantaneous velocity at individual time steps. This approach retains the coarse-to-fine nature of MeanFlow, where smaller intervals are emphasized during training to capture fine-grained details, while broader temporal dynamics are learned more gradually.

In the distillation process, the student model is guided by a flow-matching teacher model. The teacher model transforms an initial distribution \( p_0 \) (e.g., Gaussian noise) into a target distribution \( p_1 \) using a time-dependent vector field \( v(z_t, t; \theta) \). The state evolution \( z_t \) is governed by the following ordinary differential equation (ODE):

\begin{equation}
\frac{d}{dt} z_t = v(z_t, t; \theta), \quad z_0 \sim p_0, \quad z_1 \sim p_1, \quad t \in [0, 1]
\end{equation}

The teacher’s loss function is given by:

\begin{equation}
L_{\text{CFM}} = \mathbb{E}_{t, p_0(z_0), q(z_1)} \left[ \left\| v(z_t, t; \theta) - (z_1 - z_0) \right\|^2 \right]
\end{equation}

While the teacher learns to model the instantaneous velocity \( v(z_t, t; \theta) \), the student model is tasked with learning the averaged velocity over a time interval \( [t, r] \), defined as:

\begin{equation}
\bar{v}(z_t, t, r) = \frac{z_r - z_t}{r - t}
\end{equation}

where \( z_t \) and \( z_r \) are the states of the system at times \( t \) and \( r \), respectively. \( z_r \) is computed iteratively during inference as described below, and is not a direct parameter of the velocity function \( v \).

The student model is trained to approximate this averaged velocity using the teacher’s instantaneous velocity. To achieve this, we perform iterative sampling during distillation. The interval \( [t, r] \) is discretized into \( n \) subintervals, with time steps \( t_0 = t, t_1, \dots, t_n = r \). At each step, the teacher model evolves the state according to the discrete approximation of the ODE:

\begin{equation}
\label{eq:iterative-sampling}
z_{t_{k+1}} = z_{t_k} + (t_{k+1} - t_k) \cdot v(z_{t_k}, t_k; \theta)
\end{equation}

where \( t_0 = t \), \( t_n = r \), and \( t_1, t_2, \dots, t_{n-1} \) are intermediate time steps. The total displacement over the interval \( [t, r] \) is given by:

\begin{equation}
\Delta z^{\text{teacher}} = \sum_{k=0}^{n-1} \left( z_{t_{k+1}} - z_{t_k} \right) = \sum_{k=0}^{n-1} (t_{k+1} - t_k) \cdot v(z_{t_k}, t_k; \theta)
\end{equation}

This discrete displacement approximates the integral of the instantaneous velocity \( v(z_t, t; \theta) \) over \( [t, r] \), which is the continuous process the student is aiming to model. To approximate the averaged velocity, the displacement is normalized by the interval length:

\begin{equation}
\bar{v}_{\text{teacher}}(z_t, t, r) = \frac{\Delta z^{\text{teacher}}}{r - t}
\end{equation}

The continuous form of the averaged velocity is given by:

\begin{equation}
\bar{v}(z_t, t, r) = \frac{1}{r - t} \int_t^r v(z_\tau, \tau; \theta) \, d\tau
\end{equation}

Thus, the teacher’s discrete displacement serves as a numerical approximation of this integral. Finally, the student model is trained to minimize the distillation loss:

\begin{equation}
L_{\text{distill}} = \mathbb{E}_{t, r} \left[ \left\| u_{\text{student}}(z_t, t, r) - \bar{v}_{\text{teacher}}(z_t, t, r) \right\|^2 \right]
\end{equation}

where \( u_{\text{student}}(z_t, t, r) \) is the predicted velocity from the student model, and \( \bar{v}_{\text{teacher}}(z_t, t, r) \) is the target velocity from the teacher. The student model learns to predict the averaged velocity by following the teacher’s guidance, which approximates the integral of the instantaneous velocity via iterative sampling.

\begin{figure}[h]
    \centering
    \includegraphics[width=\linewidth]{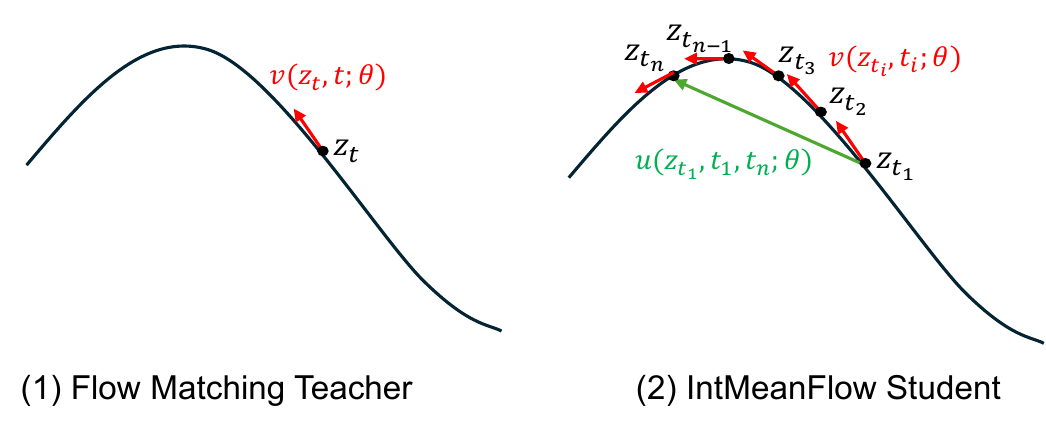}
    \caption{Illustration of IntMeanFlow: The student model learns the averaged velocity from the instantaneous velocities provided by the teacher model at multiple states.}
    \label{fig:placeholder}
\end{figure}

\vspace{-2em}

\subsection{Optimal Step Sampling Searching~(O3S)}

\begin{figure}[h]
    \centering
    \includegraphics[width=1\linewidth]{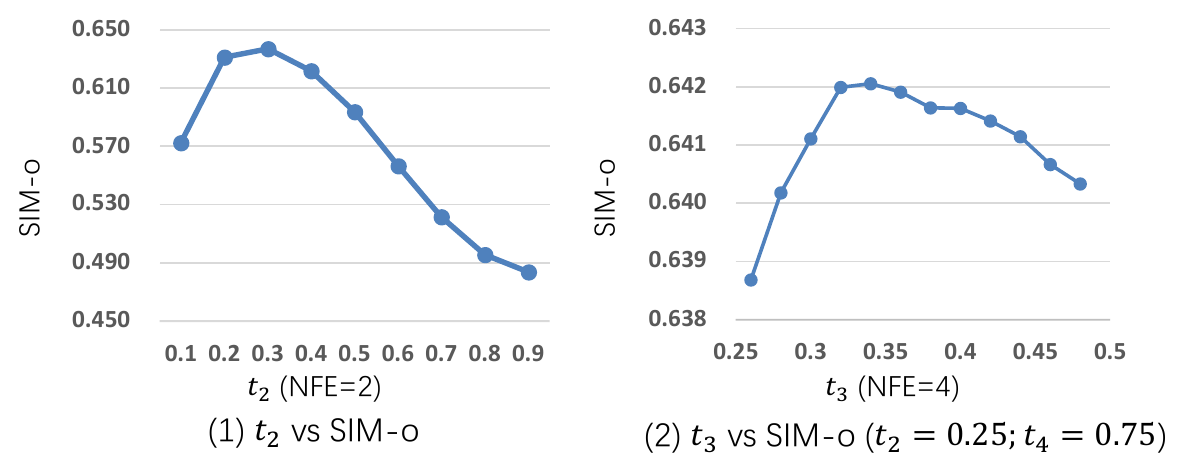}
    \caption{When fixing all but one of the sampling steps, the speaker similarity metric exhibits near-convex behavior.}
    \vspace{-1em}
    \label{fig:o3s}
\end{figure}

In flow-based speech generation, previous work has shown that denser sampling at noisier steps leads to improved speech quality~\cite{F5TTS-chen2024,EPSS-Zheng2025}. To satisfy varying NFE requirements, earlier methods have used continuous functions or hard-coded discrete step schedules. However, instead of relying on a fixed schedule, this work optimizes the steps sampling specifically for each model's inference process. As shown in Fig.~\ref{fig:o3s}, we observe through experiments that speech quality, as a function of sampling step position, exhibits near-convex behavior. This finding motivates the introduction of the Optimal Sampling Step Search (O3S) algorithm, which optimizes the distribution of a fixed number of sampling steps across the inference interval \( [0, 1] \).

The core idea behind O3S is to optimize the placement of each sampling step using ternary search. To achieve this, we fix all but one of the sampling steps and apply ternary search to optimize the remaining one. This process is repeated for each step, and the optimization continues until no further improvement is observed on a development set. O3S thus identifies the optimal distribution of sampling steps, enhancing speech quality without increasing NFE.

The pipeline for O3S is outlined in Algorithm~\ref{alg:o3s}. The metric function \( \mathcal{L} \) takes a set of sampling steps \( T \) and a development set for inference, computing a pre-defined metric on the generated audio. In this work, we use speaker similarity as the metric.

\begin{algorithm}[h]
\caption{Optimal Step Sampling Search (O3S) Algorithm}
\label{alg:o3s}
\begin{algorithmic}[1]
\State Uniformly initialize \( \mathbf{T_{best}} \gets \{t_0, t_1, \dots, t_n\} \) where \( t_i = \frac{i}{n} \) % \Comment{Initialize the sampling steps}
\State Initialize best metric \(  \mathbf{m_{\text{best}}} \gets -\infty \), patience $p \gets 0$
\State Define metric function $\mathcal{L}$

\While{True}
    \For{ \( i \in \{n-1, n-2, \dots, 1\} \)}
        \State \( m_{\text{optim}}, T_{\text{optim}} \gets \text{TernarySearch}(\mathcal{L}, i, t_{i-1}, t_{i}) \) \Comment{find best placement for $t_i$ }
         \If{ $m_{\text{optim}} > m_{\text{best}}$ } 
            \State \( T_{\text{best}}, m_{\text{best}} \gets T_{\text{optim}}, m_{\text{optim}}\)
            \State $p \gets 0$
        \Else
            \State $p \gets p+1$
            \If{ p = n}
                \State \textbf{Break}
            \EndIf
        \EndIf
    \EndFor
\EndWhile
\State \textbf{Return} $T_{\text{best}}, m_{\text{best}}$
\end{algorithmic}
\end{algorithm}
\vspace{-1em}

\subsection{Initialization Strategy for IntMeanFlow}

To adapt flow-matching models to IntMeanFlow, we introduce an additional parameter \( r \). Both \( t \) and \( r \) are passed through the same embedding network, concatenated, and projected back to the feature space of \( t \) using a linear mapping \( \mathbf{W} \).

Let the embeddings for \( t \) and \( r \) be \( \mathbf{e}_t = \mathcal{E}(t) \) and \( \mathbf{e}_r = \mathcal{E}(r) \). The concatenated and mapped embeddings $\mathbf{e}_{t,r}$, $\mathbf{e'}_{t,r}$ are:

\begin{equation}
\mathbf{e}_{t, r} = [\mathbf{e}_t, \mathbf{e}_r]; \mathbf{e'}_{t,r} = \mathbf{W} \mathbf{e}_{t, r}
\end{equation}

To preserve the original model behavior, \( \mathbf{W} \) is initialized as:

\begin{equation}
\mathbf{W} = \begin{bmatrix} D_{diag} & 0 \end{bmatrix}
\end{equation}

where $D_{diag}$ indicates a diagonal matrix. This ensures the model to behave like the original model at initialization.

\begin{table*}[ht!]
\centering
\resizebox{1\linewidth}{!}{
\begin{tabular}{llllccccccc}
\toprule
\# & \textbf{Model (NFE)} & \textbf{Data (hrs)}  & \textbf{Teacher (NFE)} & \multicolumn{1}{c}{\textbf{WER(\%)↓}} & \multicolumn{1}{c}{\textbf{SIM-o↑}} & \multicolumn{1}{c}{\textbf{UTMOS↑}}  & \multicolumn{1}{c}{\textbf{UV.MOS↑}} & \multicolumn{1}{c}{\textbf{CMOS↑}} & \multicolumn{1}{c}{\textbf{SMOS↑}} &  \multicolumn{1}{c}{\textbf{RTF↓}} \\
\midrule
\multicolumn{10}{c}{\textbf{Seed-TTS \textit{test-en}}} \\
\midrule
1 & Human & N/A & N/A & 2.14 & 0.73 & 3.52  & 3.86  & 0.00 & 3.96 & N/A \\
2 & Medium (32) & Emilia (95K)  & N/A & 1.72 & 0.70  & 3.63  & 4.03  & -0.23 & 3.92 & 0.284 \\
 3 & Base (32) & Emilia (95K)  & N/A & 1.87 &  0.67  & 3.70  & 4.06 & -0.48 & 3.88 & 0.243 \\
4 & Small (32) & LibriTTS (585)  & N/A & 2.29 &  0.58  & 3.97  & 4.16 & -0.46 & 3.23 & 0.171 \\
\cdashline{1-11}\noalign{\vskip\belowrulesep}
5 & Base + IMF (1) & Emilia (95K)  & Base (16) & 7.27 & 0.48  & 1.84  & 2.33  & - & - & 0.009 \\
 6 & Base + IMF (2) & Emilia (95K)  & Base (16) & 4.48 & 0.59  & 3.35  & 3.65  & -1.11   & 3.44 & 0.013 \\
7 & Base + IMF + O3S (2) & Emilia (95K)  & Base (16) & 2.04  & 0.63  & 3.24  & 3.58  & -0.86 & 3.52 & 0.013 \\ 
8 & Base + IMF + O3S (3) & Emilia (95K)  & Base (16) & \textbf{1.60} & \textbf{0.65}  &  \textbf{3.79}  & \underline{3.94}  & \underline{-0.61} & \textbf{3.73} & 0.021 \\
9 & Base + IMF + O3S (3) & Emilia (95K)  & Base (4) & \underline{1.71} & \underline{0.64} & \underline{3.76}  & 3.89  & -0.62 & 3.57 & 0.021 \\
10 & Base + IMF + O3S (3) & Emilia (95K)  & Base (2) & 1.83 & \underline{0.64} & 3.58  & 3.87  & -1.00 & 3.62 & 0.021 \\
11 & Small + IMF + O3S (3)  & LibriTTS (585)  & Base (16) & 1.97 & 0.63 & 3.63  & 3.89  & \textbf{-0.51} & 3.46 & 0.018 \\
 12 & Small + IMF + O3S (3) & LibriTTS (585)  & Medium (16) & 1.83 & 0.63 & 3.73  & \textbf{3.98}  &  -0.72 & \underline{3.65} & 0.018 \\
\midrule
\multicolumn{10}{c}{\textbf{Seed-TTS \textit{test-zh}}} \\
\midrule
13 & Human & N/A  & N/A & 1.25  & 0.76  & 2.78 & 4.50  & 0.00 & 3.94 & N/A \\
 14 & Base (32) & Emilia (95K)  & N/A & 1.52 & 0.76  & 2.96  & 4.56  & +0.13 & 3.87 & 0.243 \\
\cdashline{1-11}\noalign{\vskip\belowrulesep}
15 & Base + IMF (1) & Emilia (95K)  & Base (16) & 8.75 & 0.55  & 1.42  & 2.79  & - &  - & 0.009 \\
 16 & Base + IMF (2) & Emilia (95K) & Base (16) & 5.97 & 0.69  & 2.09  & 3.72  & -0.26 & 3.33 & 0.013 \\
 17 & Base + IMF + O3S (2) & Emilia (95K)  & Base (16) & 1.73 & 0.72 & 2.47  & 4.18  & \underline{0.00} & 3.65 & 0.013  \\
 18 & Base + IMF + O3S (3) & Emilia (95K)  & Base (16) & \textbf{1.67} & \textbf{0.74} &  \textbf{3.03}   & \textbf{4.48} &  \textbf{+0.07} & \textbf{3.83}  & 0.021 \\
 19 & Base + IMF + O3S (3) & Emilia (95K)  & Base (4) &  1.74 & 0.73 & \underline{2.97}  & \underline{4.45}  & -0.20  &  3.68 & 0.021 \\
20 & Base + IMF + O3S (3) & Emilia (95K)  & Base (2) & \textbf{1.67} &\textbf{0.74} & 2.84 & 4.41  & -0.29 & \underline{3.78}  & 0.021 \\
\bottomrule
\end{tabular}
}
\caption{Text2Mel results on Seed-TTS \textit{test-en} and \textit{test-zh}. Bold indicates the best result, underlined the second-best. All methods are based on F5-TTS. IMF stands for IntMeanFlow. Medium models have 592M parameters, base models 336M, and small models 158M.}
\label{tab:seedtts-test}
\vspace{-.5em}
\end{table*}

\begin{table*}[ht]
\centering
\resizebox{1\linewidth}{!}{
\begin{tabular}{llllccccccc}
\toprule
\# & \textbf{Model (NFE)} & \textbf{Data (hrs)}  & \textbf{Teacher (NFE)} & \multicolumn{1}{c}{\textbf{WER(\%)↓}} & \multicolumn{1}{c}{\textbf{SIM-o↑}} & \multicolumn{1}{c}{\textbf{UTMOS↑}} & \multicolumn{1}{c}{\textbf{UV.MOS↑}} & \multicolumn{1}{c}{\textbf{CMOS↑}} & \multicolumn{1}{c}{\textbf{SMOS↑}}  & \multicolumn{1}{c}{\textbf{RTF↓}} \\
\midrule
1 & Human & N/A &  N/A & 2.23 & 0.69 & 4.09   & 4.20 & 0.00 & 3.93 & N/A \\
2 & CosyVoice2 (32) & Proprietary (170K)   & N/A & 2.17 & 0.66 & 4.36  & 4.48  & -0.01 & 3.71 & 0.510 \\
3 & CosyVoice2 + MF (1) & LibriTTS (585)  & N/A & 2.11 & 0.62 & 3.96  & 3.85  & -0.73 & 3.42 & 0.026 \\
4 & CosyVoice2 + IMF (1) & LibriTTS (585)  & official (16) & 2.18 & 0.63 & 4.28  & 4.47  & -0.03 & 3.39 & 0.026 \\
\bottomrule
\end{tabular}
}
\caption{Token2Mel Results on LibriSpeech-PC: RTF values are reported only for the flow module.}
\vspace{-2em}
\label{tab:cosyvoice-librispeech-pc}
\end{table*}

\section{Experiments}
\label{sec:experiments}
\vspace{-1em}
\subsection{Experiment Setup}

We conduct experiments based on F5-TTS for the text2mel task and CosyVoice2 for the token2mel task. During distillation with IntMeanFlow, the student model learns the vector field from the teacher with classifier-free guidance (CFG)~\cite{CFG-ho2021}, while the student avoids using CFG during training to reduce inference overhead. For F5-TTS, we use the SeedTTS test set~\cite{seedtts-anastassiou2024}\footnote{\url{https://github.com/BytedanceSpeech/seed-tts-eval}}; for CosyVoice2, we use the LibriSpeech-PC test set~\cite{librispeech-pc-Meister2023}.

Performance is evaluated on a cross-sentence task using a variety of metrics. For objective evaluation, we report Word Error Rate (WER) and SIM-o (speaker similarity). WER is computed using Whisperlarge-v3 for English transcription and Paraformer-zh~\cite{paraformer-gao2022} for Chinese transcription. For SIM-o, we extract speaker embeddings using a WavLM-based~\cite{wavlm-chen2022} speaker verification model and compute the cosine similarity between synthesized and prompt human speech. Real-time factors(RTF) are measured on NVIDIA A100 GPU.
For CMOS, speech quality is rated from 3 (worse) to + 3 (better) compared to human reference. For SMOS, similarity between synthesized and prompt speech is rated on a 1–5 scale, with higher scores indicating better quality.
Additionally, we report UTMOS~\cite{utmos-saeki2022} scores, which are evaluated using an open-source MOS prediction model. To complement UTMOS, we also apply the recent Uni-Versa-Ext (UV.MOS)~\cite{uvmos-wang2025}, which has demonstrated a high correlation with human-provided MOS scores~\footnote{\url{https://huggingface.co/vvwangvv/universa-ext_wavlm-base_5metric}}.
\vspace{-1em}
\subsection{Results on Text2Mel Task}

\begin{figure}[h]
    \centering
    \includegraphics[width=\linewidth]{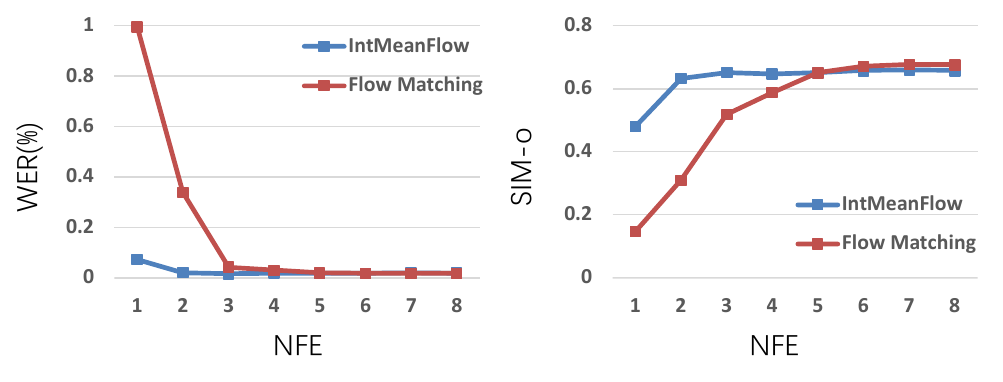}
    \caption{NFE vs. WER (\%) and SIM-o for the Flow Matching teacher and IntMeanFlow student models}
    \label{fig:nfe}
\vspace{-2em}
\end{figure}

F5-TTS maps text embeddings to mel spectrograms in the text2mel task using a diffusion transformer. Following \cite{F5TTS-chen2024}, We train small models on the LibriTTS~\cite{libritts-zen19} dataset and base and medium models on the processed 95k-hour Emilia~\cite{Emilia-he2024} dataset. Teachers' inference has a CFG rate of 3.0 while student does not apply CFG during inference.
Experimental results, shown in Table~\ref{tab:seedtts-test}, demonstrate that distillation with IntMeanFlow significantly improves NFE and RTF, with only a slight compromise in performance. The relationship between NFE and speech quality are illustrated in Fig.~\ref{fig:nfe}.
% The real-time factor (RTF) does not increase linearly with NFE, as text embedding conditions are cached across multiple NFEs. 
% Unlike IntMeanFlow students, which avoid classifier-free guidance (CFG) during inference to reduce overhead, flow-matching teachers use CFG during inference.
We do not compare with MeanFlow for the text2mel task, as F5-TTS requires training with a batch size of 1 due to JVP overhead. Performance of MeanFlow is discussed in Section~\ref{sec:token2mel}.

% FIXME: placeholder conclusion, need to confirm with objective evaluation
\subsubsection{Effectiveness of the O3S Algorithm}

The comparison between lines 6 and 7 on \textit{test-en} and lines 16 and 17 on \textit{test-zh} demonstrates the effectiveness of applying the sampling steps optimized by O3S. O3S consistently leads to improvements in both subjective and objective metrics by a large margin.

\subsubsection{Impact of Teacher NFE on Training Time and Performance}

Using a larger teacher NFE during distillation can significantly increase training time, as shown in Table~\ref{tab:train_time}. To examine the trade-off between teacher NFE and student performance, we conduct ablation experiments. A comparison of lines 10, 11, and 12 on \textit{test-en}, and lines 18, 19, and 20 on \textit{test-zh}, reveals that while a smaller teacher NFE results in minimal degradation in subjective metrics, objective metrics show noticeable performance loss, which is a common challenge in TTS evaluation. Synthesized speech from lines 10 and 20 demonstrates unnatural pronunciation and degraded fluency due to the excessively small teacher NFE.
\vspace{-1em}
% Additionally, Table~\ref{tab:train_time} presents the training time for different teacher NFE. 
% Notably, the slowdown in training time is not linearly proportional to the teacher NFE, as the teacher requires no backpropagation and its conditions are cached across multiple NFE.

\begin{table}[ht]
\centering
\begin{tabular}{lcccccc}
\toprule
Teacher NFE & N/A & 2 & 4 & 8 & 16 \\
\midrule
Time per step (seconds) & 0.93  & 1.12  & 1.74 & 2.91 & 5.17 \\
\bottomrule
\end{tabular}
\caption{Teacher NFE and training time per step}
\label{tab:train_time}
\vspace{-2.5em}
\end{table}

\subsubsection{Impact of Teacher Size on Student Performance}

The size of the teacher model does not affect the inference efficiency of the student model. Therefore, we examine whether a larger teacher can improve student performance, particularly when paired with a smaller, more practically efficient student model. Comparing lines 11 and 12, we observe that increasing the teacher's size leads to improvements in both objective and subjective metrics for the student. These improvements significantly surpass the baseline performance reported in line 4.

\vspace{-1em}
\subsection{Results on Token2Mel Task}
\vspace{-.5em}
\label{sec:token2mel}
In this experiment, we use the CosyVoice 2 architecture for flow-based speech generation. Unlike F5-TTS, where the flow operates on raw text embeddings, the flow in CosyVoice 2 processes time-aligned semantics, simplifying the task and enabling 1-NFE inference after distillation. We perform distillation using the default model configuration from the official repository\footnote{\url{https://github.com/FunAudioLLM/CosyVoice}}
 and its pre-trained checkpoint as the teacher model. Since CosyVoice 2 is trained on a proprietary dataset that is unavailable to us, we use LibriTTS for distillation. The vanilla MeanFlow model, as shown in line 18, yields good WER and SIM-o results but leads to degraded speech quality in both neural and human MOS evaluations, likely due to the absence of a teacher model and instability in the training process. The results in line 9 demonstrate that IntMeanFlow effectively distills the teacher’s capabilities while reducing NFE, even when trained on a significantly smaller dataset.

 \vspace{-1em}

\section{Conclusions}

In this work, we introduce the IntMeanFlow framework for efficient few-step speech generation through distillation. We also propose an optimal step sampling search algorithm to identify model-specific sampling steps during inference. Our experiments, conducted on two widely used TTS models, F5-TTS for the text2mel task and CosyVoice2 for the token2mel task. Results demonstrate that IntMeanFlow achieves 3-NFE for the text2mel task and 1-NFE for the token2mel task. This results in 10 times acceleration in terms of RTF, with only minimal degradation in performance. Additionally, we propose an efficient initialization strategy that facilitates the seamless adaptation of existing flow-matching models to IntMeanFlow, improving both the practical applicability of the framework.

\bibliographystyle{IEEEtran}
\bibliography{strings,refs}

\end{document}